\documentclass[aip,amsmath,amssymb,reprint,]{revtex4-1}

\usepackage{graphicx}
\usepackage{dcolumn}
\usepackage{bm}
\usepackage[utf8]{inputenc}
\usepackage[T1]{fontenc}
\usepackage{mathptmx}

\begin{document}

\preprint{AIP/123-QED}

\title{Giant resonant enhancement of optical binding of dielectric
disks}
\author{E.N. Bulgakov}
\affiliation{Kirensky Institute of Physics, Federal Research
Center KSC SB RAS, 660036 Krasnoyarsk, Russia}
\author{K.N. Pichugin}
\affiliation{Kirensky Institute of Physics, Federal Research
Center KSC SB RAS, 660036 Krasnoyarsk, Russia}
\author{A. F. Sadreev}%
 \email{almas@tnp.krasn.ru.}
\affiliation{Kirensky Institute of Physics, Federal Research
Center KSC SB RAS, 660036 Krasnoyarsk, Russia}
\date{\today}
\begin{abstract}
Two-parametric variation over the aspect ratio of each disk and
distance between disks gives rise to numerous events of avoided
crossing of resonances of individual disks. For these events the
hybridized anti-bonding resonant modes can acquire a morphology
close to the Mie resonant mode with high orbital momentum of
equivalent sphere. The $Q$ factor of such resonance can exceed the
$Q$ factor of isolated disk by two orders in magnitude. We show
that dual incoherent counter propagating coaxial Bessel beams with
power $1mW/\mu m^2$ with frequency resonant to such a anti-bonding
modes result in unprecedented optical binding forces up to decades
of nano Newtons for silicon micron size disks. We show also that a
magnitude and sign of optical forces strongly depend on the
longitudinal wave vector of the Bessel beams.
\end{abstract}

\maketitle
\section{Introduction}
The response of a microscopic dielectric object to a light field
can profoundly affect its motion. A classical example of this
influence is an optical trap, which can hold a particle in a
tightly focused light beam \cite{Ashkin1986}. When two or more
particles are present, the multiple scattering between the objects
can, under certain conditions, lead to optically bound states.
This is often referred to peculiar manifestation of optical forces
as optical binding (OB), and   it   was first   discovered   by
Burns  et   al.  on   a   system   of two  plastic   spheres   in
water  in   1989 \cite{Burns1989}. Depending on the particle
separation, OB leads to attractive or repulsive forces between the
particles and, thus, contributes to the  formation  of  stable
configurations of particles. The phenomenon of OB can be realized,
for example, in dual incoherent counter propagating beam
configurations
\cite{Tatarkova2002,Gomez-Medina2004,Metzger2006,Metzger2006a,Dholakia2010,Bowman2013,Thanopulos2017}.
Many researchers have analyzed OB force quantitatively in theory.
Chaumet {\it et al} \cite{Chaumet2001} and Ng {\it et al}
\cite{Ng2005} calculated the OB force under illumination of two
counter propagating plane waves. \u{C}i\u{z}m\'{a}r et al
\cite{Cizmar2006} presented the first theoretical and experimental
study of dielectric sub-micron particle behavior and their binding
in an optical field generated by interference of two counter
propagating Bessel beams.

An excitation of  the resonant modes with high $Q$ factor in
dielectric structures results in large enhancement of near
electromagnetic (EM) fields and respectively in extremely large EM
forces proportional to squared EM fields. First, sharp features in
the force spectrum, causing mutual attraction or repulsion between
successive photonic crystal layers of dielectric spheres under
illumination of plane wave has been considered by Antonoyiannakis
and Pendry \cite{Antonoyiannakis1997}. Because of periodicity of
structure each layer is specified by extremely narrow resonances
which transform into the sharp resonant bonding and anti-boding
resonances for close approaching of the layers. Also it was
revealed that the lower frequency bonding resonance forces act to
push the two layers together and the higher frequency anti-bonding
resonance to pull them apart. Later these disclosures we reported
for coupled photonic crystal slabs \cite{Liu09}, for coupled
asymmetric membranes \cite{Rodriguez2011}, and  two planar
dielectric photonic metamaterials \cite{Zhang2014} due to
existence of resonant states with infinite $Q$ factor (bound
states in the continuum).

\begin{figure}
\includegraphics*[width=0.8\linewidth]{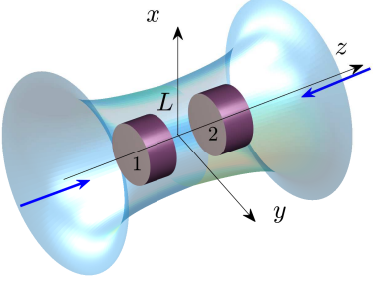}
\caption{Two silicon disks with the radius $a$, permittivity
$\epsilon=15$ are illuminated by dual counter-propagating mutually
incoherent Bessel beams with zero azimuthal index $m=0$. Light
intensity of each beam $1mW/\mu m^2$.} \label{fig1}
\end{figure}

Even two particles can demonstrate precedents of extremely high
$Q$ factors resonant modes owing to the mechanism of avoided
crossing. The bright example is avoided crossing of
whispering-gallery modes (WGM) in coupled microdisks
\cite{Benyoucef2011} which resulted in extremely high enhancement
of OB between coupled WGM spherical resonators
\cite{Povinelli2005}. However the WGM modes can be excited only in
spheres with large radii of order $30\mu m$. Respectively, the OB
force for such massive particles can be turned out not so
significant.  In the present letter we offer a solution to the
problem by use of two coaxial silicon disks of micron sizes shown
in Fig. \ref{fig1} which show extremely high-$Q$ factors in the
subwavelength regime. Owing to two-parametric (over the aspect
ratio and distance between disks) avoided crossing of low order
resonances the anti-bonding resonant mode acquires a morphology of
the high order Mie resonant mode of effective larger sphere with
extremely small radiation losses \cite{Bulgakov2020}. We show also
that a magnitude and what is more interesting the sign of the OB
force strongly depend on the wave number of the Bessel beams that
opens additional options to manipulate high index particles
optically.

\section{Two-parametric avoided crossing of resonances}
\begin{figure}[ht!]
\includegraphics[width=0.9\linewidth]{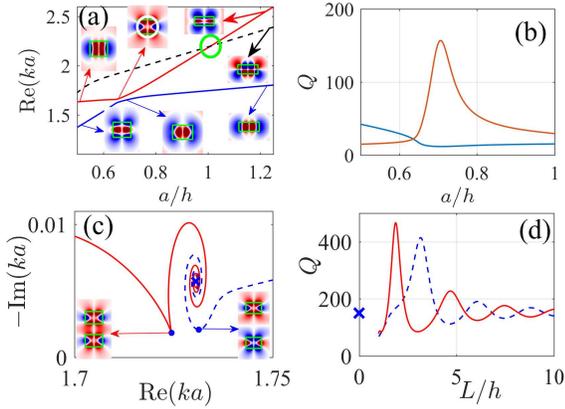}
\caption{(a) Avoided crossing of two TE resonances whose modes are
symmetric relative to $z\rightarrow -z$ and (b) their $Q$ factors
versus the aspect ratio $a/h$ in isolated silicon disk with
dielectric constant $\epsilon=12$. Insets show the profiles of
tangential component of electric field $E_{\phi}$. Crosses mark
the degenerate resonant frequencies and respectively the $Q$
factor of isolated disks. (c) and (d) Evolution of resonances and
the $Q$ factor vs distance between disks at $a/h=0.71$.}
\label{fig2}
\end{figure}

The phenomenon of avoided resonance crossing has attracted
interest in photonics by the enhancement of the quality factor of
resonant modes in coupled optical microcavities in the WGM regime
\cite{Wiersig2006,Boriskina2006}. An interest is renewed when high
$Q$ resonant modes were revealed even in isolated dielectric disk
show high $Q$ factor owing to avoided crossing of resonant modes
for variation of aspect ratio in {\it subwavelength} range
\cite{Rybin2017}. Because of importance of this result we
reproduce that process in Fig. \ref{fig2} (a).
At the aspect ratio $a/h=0.71$ the $Q$ factor reaches maximal
value around 160 for high index optical material (Si) as shown in
Fig. \ref{fig2} (b). One can see that the hybridized mode has a
morphology very close to the morphology of the Mie resonant mode
with orbital momentum $l=3$ in equivalent sphere whose volume
equals $\pi a^2h$. This mode highlighted by white circle in upper
inset of Fig. \ref{fig2} (a). The spherical particle has the
minimal surface compared to given volume and therefore smallest
radiation losses compared to particles with another shapes. That
explains a reason to enhance the $Q$ factor by one order compared
to the disk.
That explains the peak in the $Q$ factor of the corresponding
resonant mode in Fig. \ref{fig2} (b).

If to fix this optimal aspect ratio 0.71 and traverse over the
distance between disks the $Q$ factor enhances by a few times
compared to the isolated disk \cite{Pichugin2019}. Evolution of
the resonances and the behavior of the quality factor
$Q=-\frac{{\rm Re}(ka)}{2{\rm Im}(ka)}$ are shown in Fig.
\ref{fig2} (c) and (d) respectively. At $L\gg a$ the resonances
are degenerate that is marked by cross in Fig. \ref{fig2} (c). For
an approaching of disks the resonances are split and evolve
spirally so that at some distances the imaginary parts of
hybridized complex resonant frequencies reach some minima marked
in Fig. \ref{fig2} (c) by closed circles. The spiral behavior of
complex resonant frequencies is a result of radiation of leaky
resonant modes by one disk and consequent scattering by the
another. These scattering processes  give rise to the coupling
$e^{ikL}/L^2$ which hybridizes the resonant modes of separate
disks as the leaky bonding and anti-bonding resonant modes
\cite{Pichugin2019}. They are shown in insets of Fig. \ref{fig2}
(c) at left (bonding mode) and right (anti-bonding mode)  at those
distances at which the $Q$ factor reaches the maxima, $L/a=1.87$
and $L/a=3.16$ respectively as shown in Fig. \ref{fig2} (d).

\section{Optical binding forces}
It is clear that the same mechanism of consequent scattering
processes underlines the optical binding stimulated by incident
Bessel beam and respectively the OB force. It is reasonable to
consider the OB force at the optimized aspect ratio $a/h=0.71$
where the $Q$ factor of the bonding and anti-bonding resonant
modes show the maximal $Q$ factors. We consider the Bessel beams
with TE polarization in the simplest form with zero azimuthal
index $m=0$ \cite{Karasek2009}
\begin{equation}\label{Bessel}
    {\bf E}_{inc}(r,\phi,z)=E_0{\bf e}_{\phi}\exp(\pm ik_zz)J_1(k_rr)
\end{equation}
where $J_1$ is an Bessel function, $k_z$ and $k_r$ are the
longitudinal and transverse wave numbers, with the frequency
$\omega/c=k=\sqrt{k_r^2+k_z^2}$ and $r, \phi$, and $z$ are the
cylindrical coordinates, ${\bf e}_{\phi}$ is the unit vector of
the polarization. The total system of two disks and applied Bessel
beams preserves the axial symmetry that allows us to consider the
simplest case with zero azimuthal index $m=0$. In order to
stabilize both disks in z-direction we use the approach in which
two counter-propagating mutually incoherent Bessel beams were
applied \cite{Tatarkova2002,Metzger2006} that is schematically
shown in Fig. \ref{fig1}.

\begin{figure}
\includegraphics*[width=0.9\linewidth]{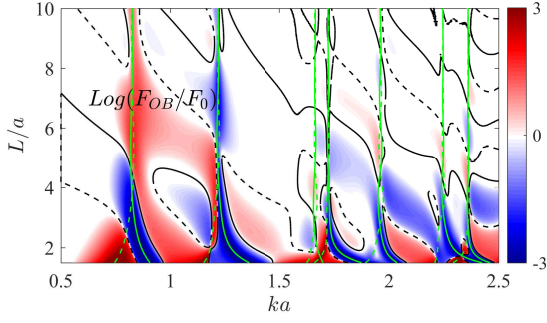}
\caption{The OB force between two disks vs the frequency and
distance between centers of disks with radius $a=0.5\mu m$, aspect
ratio $a/h=0.71$ and $\epsilon=12$ under illumination of the
Bessel beam with TE polarization and $k_za=1$. Black solid (dash)
lines show stable (unstable) configuration of disks. Green solid
(dash) lines show anti-bonding (anti symmetric) and bonding
(symmetric) resonant frequencies of two disks vs the distance
between. $F_0=1pN$. } \label{fig3}
\end{figure}

At first, we considered a stability of single disk  at $r=0$.
Numerical calculations of difference between force produced  by
the centered Bessel beam and the force produced by slightly
shifted beam show that the Bessel beams strongly trap spherical
particles at the symmetry axis, i.e., at $r=0$ (stable zero-force
points) similar to the case of sphere \cite{Milne2007}. That
considerably simplifies the further calculation of the OB force
between two disks and allows to consider the optical forces over
the axis of symmetry only. We define the OB force
$F_{OB}^{\rightarrow}=(F_{1z}-F_{2z})/2$  where the indices 1 and
2 note the disks where the Bessel beam incident at the left. Owing
to an incoherence of the Bessel beam illuminated from the right we
have the same expression for
$F_{OB}^{\rightarrow}=-F_{OB}^{\leftarrow}$. As a result we obtain
doubled value for the OB force $F_{OB}=F_{1z}-F_{2z}$ and zero
optical pressure on both disks. From Fig. \ref{fig3} one can see
that the OB force is sensitive to the resonant frequencies shown
by green solid (bonding) and dash (anti-bonding) lines.
One can that the bonding and anti-bonding resonances shown in Fig.
\ref{fig2} which achieve the $Q$ factor above 400 have no
significant bright effect effect on the OB force as Fig.
\ref{fig3} shows. We see that maximal OB force is enhanced by
three orders and reached a value up to one femto Newton.

\begin{figure}[ht!]
\includegraphics[width=1\linewidth]{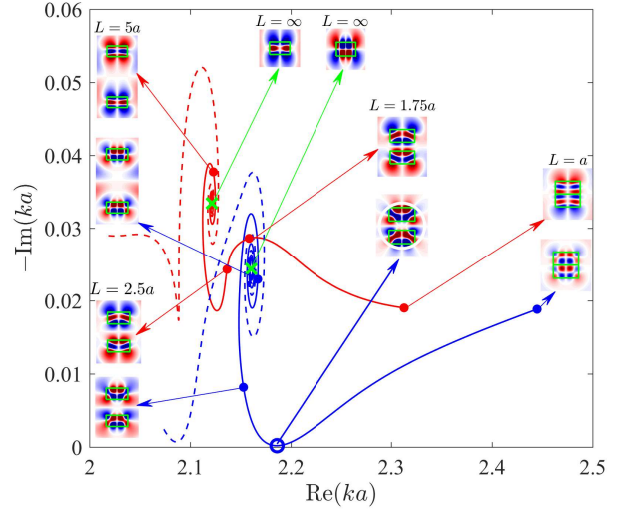}
\caption{Avoided crossing of resonances originated from coupling
of orthogonal resonances of isolated disk shown in Fig. \ref{fig2}
(a) for variation of distance between disks for $h/a=1.009$.
Solid/dash lines show the anti-bonding/bonding resonances. Insets
show the profiles of tangential component of electric field
$E_{\phi}$. } \label{fig4}
\end{figure}

\begin{figure}[ht!]
\includegraphics[width=0.9\linewidth]{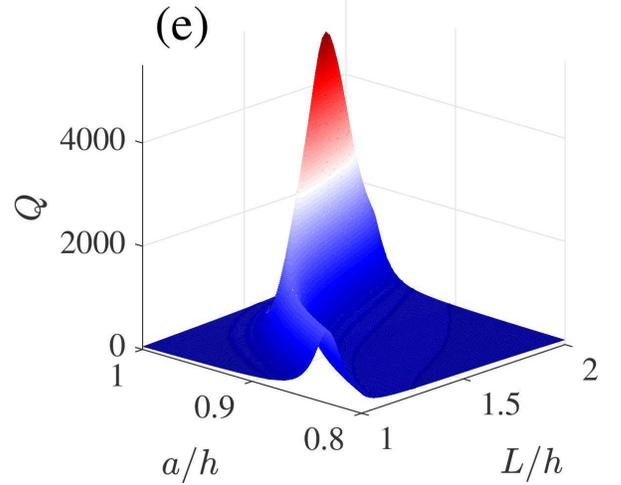}
\caption{The $Q$ factor vs aspect ratio and distance between disks
of the anti-bonding resonant mode (blue line in Fig. \ref{fig4}
highlighted by open circle).} \label{fig5}
\end{figure}

Next, we show in Fig. \ref{fig2} (a) the antisymmetric resonance
(black dash line) crosses the symmetric resonance (red solid line)
at $a/h=1.009$. These resonances are not coupled in the isolated
disk because of their orthogonality to each other. However as soon
as the second disk is approaching this symmetry prohibition is
lifted. Evolution of these resonances with distance $L$ between
disks is shown in Fig. \ref{fig4}. When the distance is large
enough the resonances marked by green crosses are degenerate. Let
define the corresponding modes as $\psi_1(\vec{r})$ and
$\psi_2(\vec{r})$ which are shown in Fig. \ref{fig4} in upper
insets. With approaching of disks these resonant modes are
hybridizing as follows \cite{Pichugin2019}
\begin{equation}\label{s,a}
    \psi_{1,2;s,a}(\vec{r})\approx\psi_{1,2}(\vec{r}_{\perp},z-\frac{1}{2}L\vec{z})
    \pm \psi_{1,2}(\vec{r}_{\perp},z+\frac{1}{2}L\vec{z})
\end{equation}
where $\vec{z}$ is the unit vector along the z-axis. These modes
can be classified as bonding (symmetrical) or anti-bonding
(anti-symmetrical) resonant modes and illustrated in Fig.
\ref{fig4} at $L=5a$. However with further approaching of disks
the approximation (\ref{s,a}) ceases to be correct because of
interaction of the resonances $\psi_1$ and $\psi_2$. One can
observe noticeable deformation of these resonant modes at $L=2.5a$
in Fig. \ref{fig4} and especially at $L=1.75a$. At this distance
and the aspect ratio $a/h=1.009$ the anti-bonding resonant mode
highlighted in Fig. \ref{fig4} by open circle is featured by
extremal low resonant width as marked by rhombus. Respectively,
one can observe extremely  high peak of the $Q$ factor around 5500
in Fig. \ref{fig5} at the vicinity $L=1.75a$ and $a/h=1.009$. The
further approaching of disks until they stick each other at $L=a$
results in the resonant modes surprisingly close to approximation
(\ref{s,a}) as insets at the right of Fig. \ref{fig4} show. The
reason of extremely small radiation losses of the anti-bonding
mode at $L=1.75a$ and $a/h=1.009$ is related to its morphology
which is immensely close to the morphology of the Mie resonant
mode with orbital momentum $l=6$ of a sphere with volume
$\pi(h+L)a^2$. That sphere is highlighted in the corresponding
inset in Fig. \ref{fig4} by open white circle.

\begin{figure}[ht!]
\includegraphics[width=1.35\linewidth]{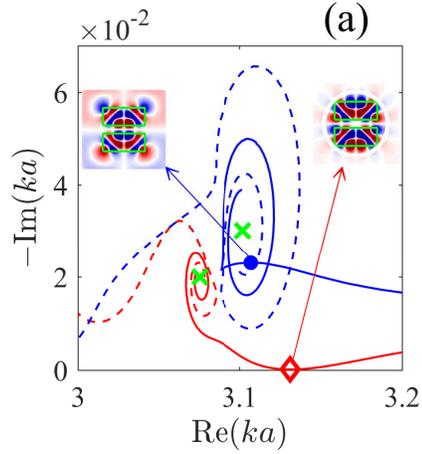}
\includegraphics[width=1.2\linewidth]{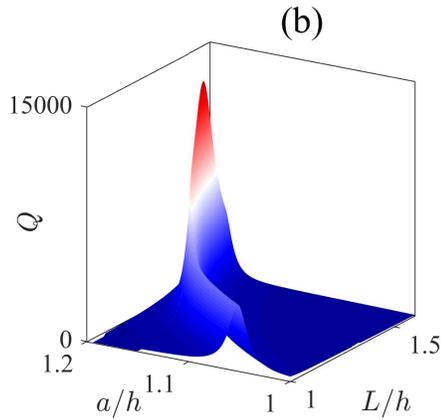}
\caption{(a) Evolution of the higher lying TE resonances in
traversing with the distance between the disks for $a/h=1.17$.
Solid/dash lines show the anti-bonding/bonding resonances. (b) The
$Q$ factor vs the distance between the disks and their aspect
ratio.} \label{fig6}
\end{figure}

It is worthy to note that this case of extremal enhancement of the
$Q$ factor due to avoided crossing of orthogonal resonant modes of
isolated disk is not unique. Fig. \ref{fig6}, for example,
demonstrates another scenario of the avoided crossing for
approaching of disks with aspect ratio $a/h=1.17$ however for
higher lying resonant modes. The right inset shows that the
anti-bonding resonant mode  which demonstrates unprecedented $Q$
factor 15000 at $L=1.4a$. Similar to the case shown Fig.
\ref{fig4} the reason of that is the morphology of the
anti-bonding resonant mode is close to the Mie resonant mode with
$l=8$.

\begin{figure}
\includegraphics*[width=1\linewidth]{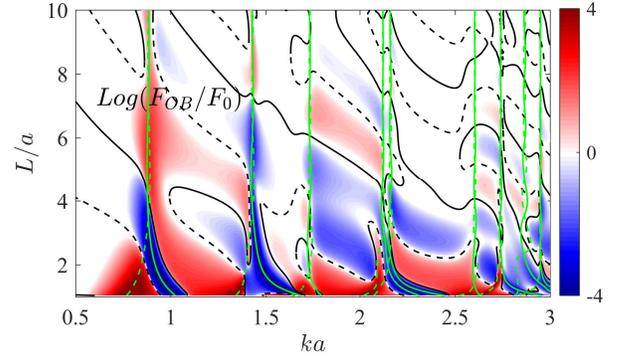}
\caption{The binding force between two disks vs the frequency and
distance between centers of disks with aspect ratio $a/h=1.0395$
for the Bessel beam with TE polarization $k_za=1/2$ where the disk
with $\epsilon=12$ has the radius $a=0.5\mu m$. } \label{fig7}
\end{figure}

Fig. \ref{fig7} shows general picture for the OB force at
$a/h=1.009$ in Log scale versus the frequency of the dual Bessel
beams and distance $L$.
\begin{figure}
\includegraphics*[width=0.9\linewidth]{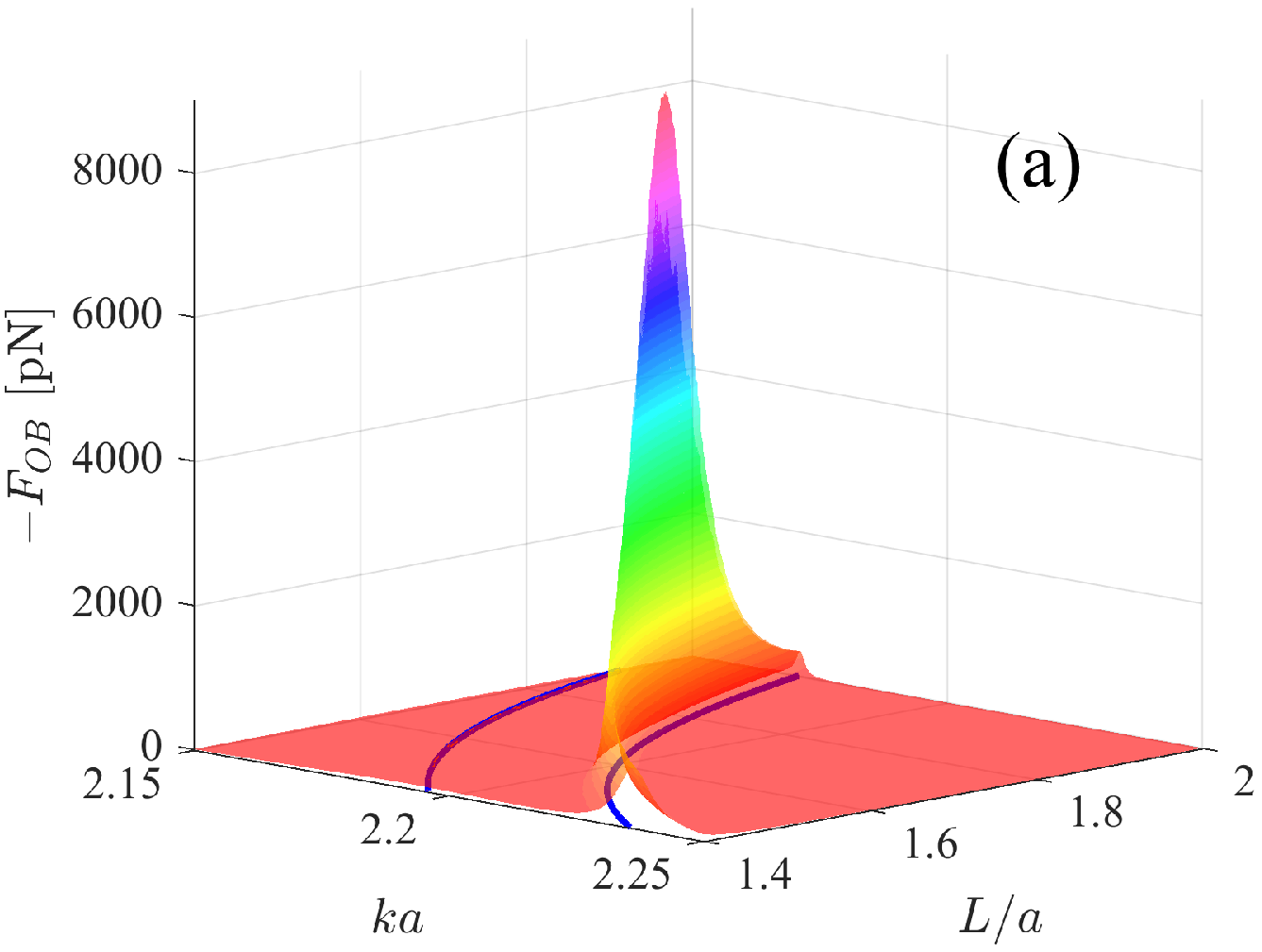}
\includegraphics*[width=0.9\linewidth]{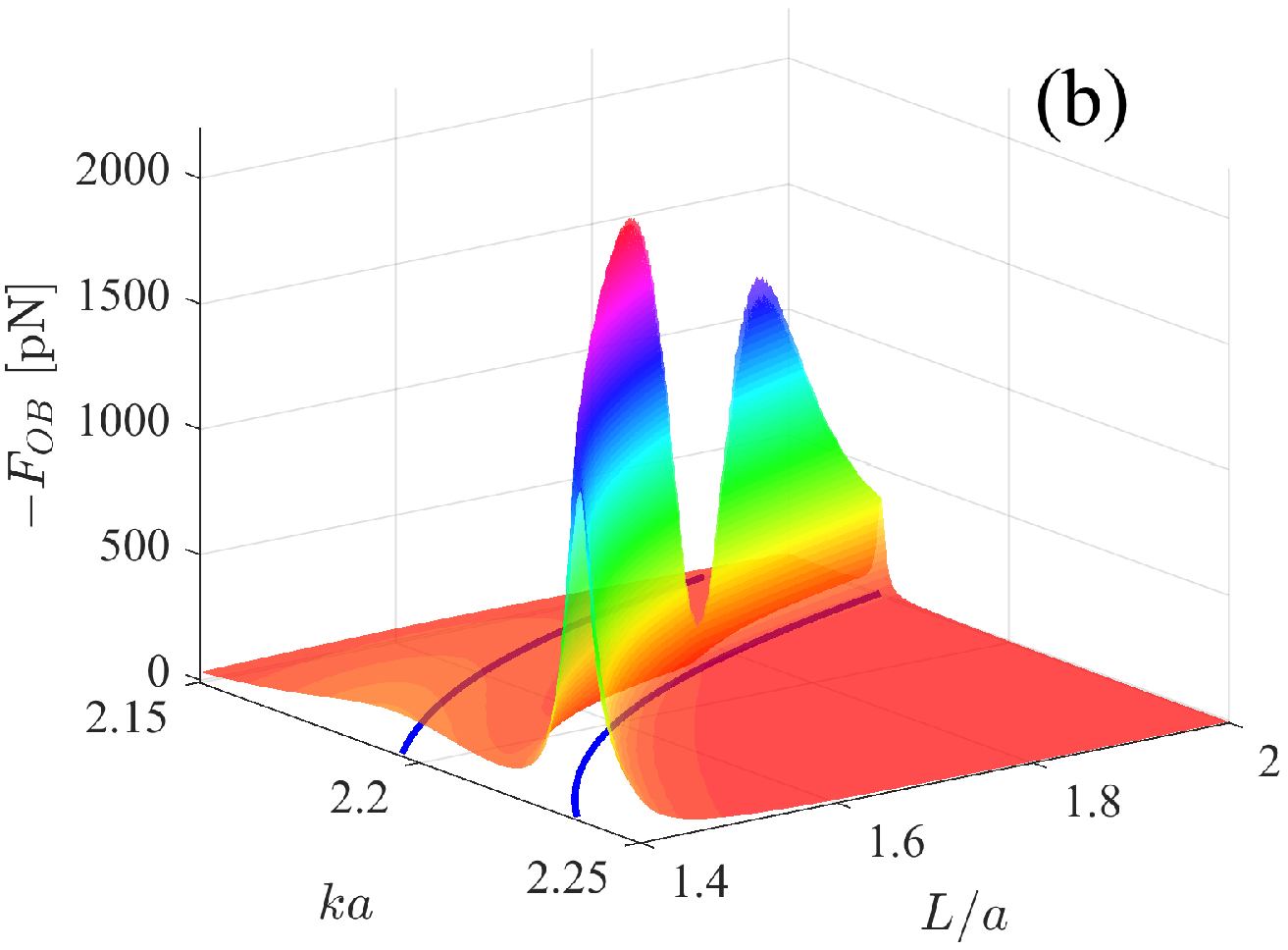}
\caption{The OB vs distance between centers of disks at the
vicinity of the anti-bonding  resonance marked in Fig. \ref{fig2}
(c) by closed circle $ka=1.95$ (a) $k_za=0.5$ and (b) $k_za=1$.
Solid line underneath shows anti-bonding resonant frequencies vs
distance $L$ highlighted in Fig. \ref{fig5}.} \label{fig8}
\end{figure}
In order the reader could imagine the extremal behavior of the OB
force we reproduce fragment of Fig. \ref{fig7} as surface in Fig.
\ref{fig8} (a) where one can see that giant OB is achieved around
30 femto Newtons at $ka=1.97, L=1.85a, h=1.03a, k_za=0.5$. Fig.
\ref{fig8} (b) shows that this giant peak is split for $k_za=1$.
It is remarkable that the equilibrium distances between disks is
traversed close to the anti-bonding resonance shown by solid line.
Fig. \ref{fig9} demonstrates as this giant peaks in OB are easily
manipulated by small changes in parameters of the Bessel beam:
$k_za$ and frequency.
\begin{figure}
\includegraphics*[width=0.9\linewidth]{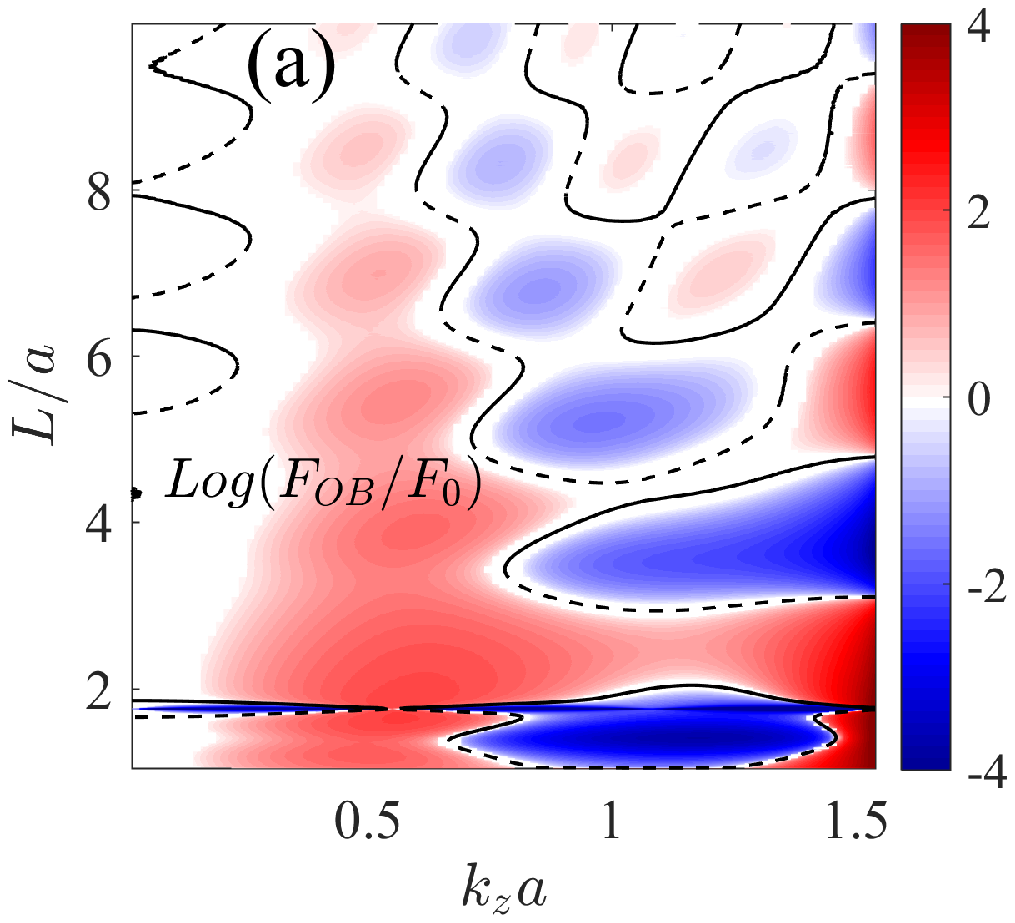}
\includegraphics*[width=0.9\linewidth]{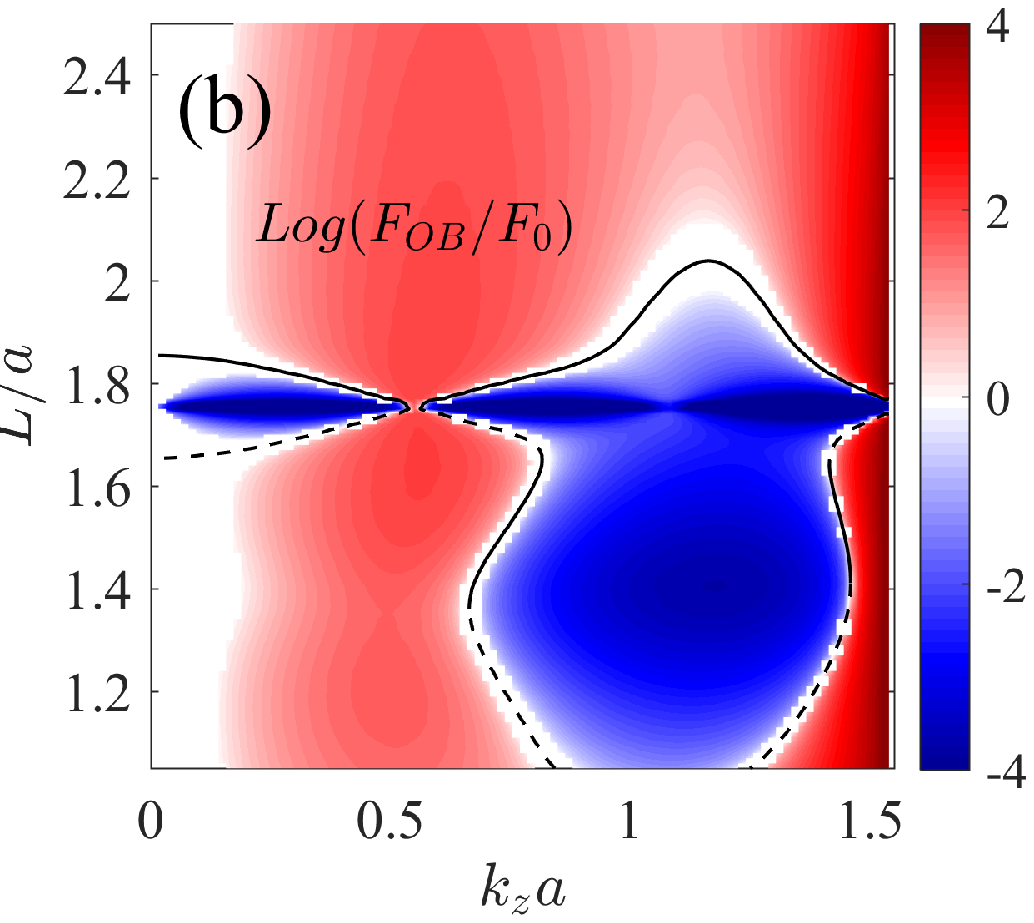}
\caption{(a) The OB vs distance between disks and longitudinal
wave vector of the Bessel beam $k_za$ at the vicinity of the
anti-bonding resonance marked in Fig. \ref{fig2} (c) by closed
circle $ka=1.95$. (b) Zoomed version of (a).} \label{fig9}
\end{figure}


\section{Summary}

In the present  paper we considered the resonant enhancement of
the OB force of two silicon disks of micron size by illumination
of dual incoherent counter propagating Bessel beams. As distinct
from the case of two dielectric spheres \cite{Chaumet2001,Ng2005}
the case of coaxial disks brings new aspect for the OB force
related to   the  extremely high $Q$ factor due to two-parametric
  avoided crossing of orthogonal resonances over aspect ratio and distance between the disks
  \cite{Bulgakov2020}. The corresponding anti-bonding resonant
modes of two disks are turned out to be closed to the Mie resonant
mode with high orbital index $l=6$ or even $l=8$ of an effective
sphere with volume $4\pi R^3/3=\pi(h+L)a^2$ \cite{Bulgakov2020}
that explains the extremely high $Q$ factors. For the case of two
coaxial silicon disks with micron diameter illuminated by dual
coaxial Bessel beams we demonstrate giant OB force in few decades
of femto Newtons in a vicinity of anti-bonding resonances.  Giant
enhancement of optical forces have been reported already
\cite{Antonoyiannakis1997,Liu09,Zhang2014} however for PhC layers.
It is remarkable that the OB force can be easily manipulate by the
counter propagating Bessel beams, by tuning of the frequency onto
the the resonances of anti-bonding resonances and by width of the
Bessel beams.

{\bf Acknowledgments}\\ The work was supported by Russian
Foundation for Basic Research projects No. 19-02-00055.

\end{document}